\documentclass[10pt]{article}

\usepackage{amsmath}
\usepackage{amssymb}

\usepackage{graphicx}

\usepackage{cite}

\usepackage{color}

\usepackage[labelfont=bf,labelsep=period,justification=raggedright]{caption}
\def\DI{{\mathrm {DI}}}
\makeatletter
\renewcommand{\@biblabel}[1]{\quad#1.}
\makeatother

\date{}

\begin{document}

\begin{flushleft}
{\Large \textbf{EEG Spatial Decoding and Classification with Logit
Shrinkage Regularized Directed Information Assessment (L-SODA)} }
\\
Xu Chen$^{1,\ast}$, Zeeshan Syed$^{1}$, Alfred Hero$^{1}$
\\
\bf{1}  Dept of EECS, University of Michigan at Ann Arbor, Ann
Arbor, MI, USA
\\

$\ast$ E-mail: Corresponding xhen@umich.edu
\end{flushleft}

\section*{Abstract}
There is an increasing interest in studying the neural interaction
mechanisms behind patterns of cognitive brain activity. This paper
proposes a new approach to infer such interaction mechanisms from
electroencephalographic (EEG) data using a new estimator of directed
information (DI) called logit shrinkage optimized directed
information assessment (L-SODA). Unlike previous directed
information measures applied to neural decoding, L-SODA uses
shrinkage regularization on multinomial logistic regression to deal
with the high dimensionality of multi-channel EEG signals and the
small sizes of many real-world datasets. It is designed to make few
a priori assumptions and can handle both non-linear and non-Gaussian
flows among electrodes. Our L-SODA estimator of the DI is
accompanied by robust statistical confidence intervals on the true
DI that make it especially suitable for hypothesis testing on the
information flow patterns. We evaluate our work in the context of
two different problems where interaction localization is used to
determine highly interactive areas for EEG signals spatially and
temporally. First, by mapping the areas that have high DI into
Brodmann area, we identify that the areas with high DI are
associated with motor-related functions. We demonstrate that L-SODA
provides better accuracy for neural decoding of EEG signals as
compared to several state-of-the-art approaches on the Brain
Computer Interface (BCI) EEG motor activity dataset. Second, the
proposed L-SODA estimator is evaluated on the CHB-MIT Scalp EEG
database. We demonstrate that compared to the state-of-the-art
approaches, the proposed method provides better performance in
detecting the epileptic seizure.

\section{Introduction}
An extensive body of research focuses on the goal of identifying and
classifying brain activity using electroencephalographic (EEG) data.
Central to these efforts is developing an understanding of how the
brain processes information to achieve specific tasks. Previous work
([1][2][6][7][11][19][20][21][26]-[32]) has shown that certain
regions in the human brain have strong interactions. As introduced
by Granger [15][27], the relation between two signals may be
expressed in terms of the linear predictability of one signal by the
knowledge of the immediate past of the other signal. The original
Granger causality measure was restricted to stationary Gaussian time
series but later versions relaxed this stationary assumption.
However, the representation and detection of the information flow
within the brain remains a challenging problem due to assumptions of
linearity and Gaussianity in measurable brain signals..

Directed information (DI) provides a measure of information that is
suitable for non-linear and non-Gaussian dependencies between
different EEG signal sources. The DI was first proposed by Massey in
1990 as an extension of Shannon's mutual information (MI) [16].
Different from MI, DI is an asymmetric function of the
time-aggregated feature densities extracted from pairs of
measurement sites [10]. In [26], Quinn et al. utilized unregularized
DI to capture the non-linear and non-Guassian dependency structure
of spike train recordings, where they estimated the DI with point
process generalized linear models. In this case, parameter and model
selection was performed using maximum likelihood estimates and
minimum description length. However, while the use of directed
information in this manner has been demonstrated to be superior to
Granger's measure and MI, due to the high dimensionality of the
features and small sample size intrinsic to EEG signals, a direct
implementation of an empirical DI estimator suffers from severe
overfitting errors. In this paper, we introduce an improved estimate
of DI, called logit shrinkage optimized directed information
assessment (L-SODA). L-SODA is conceptually simple, is of low
implementation complexity, and is mean-square optimal over the class
of regularized directed information estimators. The main difference
between our work and [26] lies in the fact that the proposed L-SODA
approach to DI estimation controls overfitting errors using
shrinkage regularization applied to a reduced set of the features of
EEG signals. This work builds upon our previous efforts [37] on
shrinkage optimized directed information assessment (SODA), and
seeks to explicitly address the curse of dimensionality by applying
shrinkage methods on multinomial logistic regression [41] to
estimate likelihood functions required for evaluating the DI. Our
experiments demonstrate that L-SODA's performance is superior to
state-of-the-art methods for neural interaction detection and
classification.

There is an extensive literature related to EEG signal interaction
detection and classification. The authors of [3] used MI to classify
EEG data. Maximum mutual information is applied to feature selection
for EEG signal classification [4]. In [2][31], a directed transfer
function (DTF) with a linear autoregressive model has been employed
to fit the data and interaction is deduced and characterized. In
[23], Hesse et al. presented an adaptive estimation of Granger
causality. Simulations demonstrate the usefulness of the
time-variant Granger causality for detecting dynamic causal
relations. In [32], Lotte et al. reviewed classification algorithms
used to design Brain Computer Interface (BCI) systems based on EEG.
In [27], Supp et al. identified the directionality of oscillatory
brain interactions in source space during human object recognition
and suggested that familiar, but not unfamiliar, objects engage
widespread reciprocal information flow. In [28], Hinrichs et al.
analyzed the timing and direction of information flow between
striate (S) and extrastriate (ES) cortex by applying a generalized
mutual information measure during a visual spatial attention task.
In [29], Babiloni et al. presented advanced methods for the
estimation of cortical connectivity from combined high-resolution
EEG and functional magnetic resonance imaging (fMRI) data.

Unlike these previous efforts, L-SODA implements estimates of the
conditional distributions of the features over each temporal segment
given the features in previous segments. This is effectively a
Markov model if a window of fixed width is used. Our L-SODA approach
is completely data-driven. It relies solely on a non-parametric
regularized estimate of the conditional feature probability
distribution. To estimate these high dimensional conditional feature
distributions, L-SODA implements a novel James-Stein shrinkage
regularization method ensuring minimum mean-squared error (MSE) over
the family of shrinkage estimators of DI. Such a shrinkage approach
was adopted by Hauser and Strimmer [5] for entropy estimation, but
without logistic models. The L-SODA approach to estimating DI used
shrinkage regularization that minimizes estimator mean square error
and provides asymptotic expressions for estimator bias, variance in
addition to a central limit theorem (CLT). Here we apply L-SODA to
obtain an empirical directed graph of interactions between EEG
electrodes, using the CLT to specify p-values for testing the
statistical significance of detected interactions via simulation and
to control the false discovery rate on the putative edge discoveries
in the graph.

To evaluate our work, we use two approaches. The first utilizes
ground truth neural functional area locations (Brodmann areas) for
validation of our localization results. The Brodmann areas were
originally defined and numbered by Korbinian Brodmann based on the
cytoarchitecture organization of neurons he observed in the cerebral
cortex [43]. When applied to a brain computer interface (BCI) motor
activity dataset, we show that the directed information graph
discovered by L-SODA is consistent with activation of the known
Brodmann areas of the brain associated with motor functions. Our
second evaluation approach assesses the utility of the interactions
discovered by L-SODA to improve a neural classification task.
Specifically, for the CHB-MIT Scalp EEG dataset, we show that L-SODA
has better ability to detect epileptic seizure onset.

For both these cases, L-SODA exhibits performance advantages over
the unregularized DI estimation of Quinn et al [26]. Since L-SODA is
specifically designed for low sample sizes, L-SODA can be viewed as
an optimized shrinkage regularized estimator of directed
information. Compared to other forms of regularization such as
sparse representation with l1 regularization [41], the error of the
logit shrinkage regularization approach can be analyzed and
optimized in the mean square sense. The shrinkage regularization of
L-SODA has significant performance gains in the low sample size
regime, a regime that is typical for temporally windowed BCI data.
Moreover, as described in [32], the features in BCI and CHB-MIT
Scalp EEG databases are non-stationary since EEG signals may vary
rapidly. We demonstrate by experiment that the proposed L-SODA
algorithm is not only able to control false positive rate more
accurately, but also has lower false negative rate in detecting
significant information flow at given false positives level.

\subsection{Summary of the main contribution}
In this paper, we:
\begin{enumerate}
\item
Propose a new approach (L-SODA) to infer neural interaction
mechanisms from EEG signals using a novel estimator of DI with logit
shrinkage optimized DI assessment.
\item
Derive a central limit theorem for this novel estimator that can be used to assess statistical significance for interaction detection.
\item
Illustrate that L-SODA can find interactions between anatomical
regions of the brain that are plausible in the context of mapping to
Brodmann functional areas [25] for different tasks.
\end{enumerate}
We further demonstrate the superiority of this approach over existing methods for DI estimation using the BCI project dataset and the CHB-MIT Scalp EEG database:
\begin{enumerate}
\item
In terms of sensitivity, as compared to unregularized DI, Granger's
measure and coherence measure, L-SODA is capable of detecting new
interactions among EEG signal sources and has at least 5\% better
localization accuracy. The localization results are verified using
neural pathway locations of motor activities and are not discovered
by existing approaches when subjected to the same false discovery
rate.
\item
In terms of specificity, as compared to previous results based on
unregularized DI estimation [26], L-SODA is better able to control
false positive rate (type I error) while maintaining high
interaction detection accuracy.
\item
L-SODA improves classification accuracy by 6\% relative to the
performance of Hidden Markov Models (HMMs), mutual information (MI),
Granger's measure, coherence measure and unregularized DI
estimation.
\item
We demonstrate that compared to unregularized DI
estimation [26], L-SODA has an 8\% lower false negative rate (type II
error) in detecting information flow at given level of false
positives.
\item
Moreover, by applying L-SODA with shrinkage logistic regression, we
reduce the number of false positives for seizure detection  by 3\%
compared to energy-based method and unregularized DI.
\end{enumerate}
The rest of the paper is organized as follows: In Section 2, we
present the statistical framework of L-SODA. In Section 3, we
subsequently propose the L-SODA-based interaction detection
algorithm. In Section 4, we evaluate the proposed algorithm and
compare its performance with the state-of-the-art approaches. We
conclude with a brief summary in Section 5.

\section{L-SODA Framework for EEG}
The SODA framework for directed information estimation was
introduced in [37] for audio-video indexing. SODA is not scalable to
high dimensional feature space since it requires discretization over
the feature space in order to apply the multinomial model of [37].
To overcome this problem, here we augment this approach with the use
of logit shrinkage (i.e., L-SODA) for use within the high
dimensional EEG context.

The EEG is an aggregate measure of neurological activity within the
brain. Consider two EEG electrodes $E_x$ and $E_y$ placed at
positions $x$ and $y$ with $M_x$ and $M_y$ time points respectively.
We denote by \textbf{$X_m$} and \textbf{$Y_m$} the temporal feature
variables extracted at time $m$ for $E_x$ and $E_y$, and define
$X^{(m)}=\{X_k\}_{k=1}^m$ and $Y^{(m)}=\{Y_k\}_{k=1}^m$. The mutual
information (MI) between $E_x$ and $E_y$ is given by $ {\mathrm
{MI}}(E_x;E_y)= \textbf{E}\left[ \ln
\frac{\textbf{f}(\textbf{X}^{(M_x)},\textbf{Y}^{(M_y)})}{\textbf{f}(\textbf{X}^{(M_x)})\textbf{f}(\textbf{Y}^{(M_y)})}\right],
$ where $f(X^{(M_x)},Y^{(M_y)})$ is a joint distribution and
$\textbf{f}(X^{(M_x)})$ and $\textbf{f}(Y^{(M_y)})$ are marginal
distributions. The DI from electrode $E_x$ to electrode $E_y$ is a
non-symmetric generalization of the MI defined as
\begin{eqnarray}
\DI(E_x\Rightarrow E_y)&=&\sum_{m=1}^{M}I(X^{(m)};Y_{m}|Y^{(m-1)})~,
\label{DIformulationp}
\end{eqnarray}
where $M= \min\{M_x,M_y\}$,
$I(X^{(m)};Y_{m}|Y^{(m-1)})$=$\textbf{E}[\ln\frac{\textbf{f}(X^{(m)};Y_{m}|Y^{(m-1)})}{\textbf{f}(X^{(m)})\textbf{f}(Y_{m}|Y^{(m-1)})}]$
is the conditional MI between $X^{(m)}$ and $Y_{m}$ given the past
$Y^{(m-1)}$. An equivalent representation of DI
 is in terms of the conditional entropies
\begin{eqnarray}
&&H(X^{(m)}|Y^{(m-1)})=\textbf{E}[\ln
\textbf{f}(X^{(m)}|Y^{(m-1)})], H(X^{(m)}|Y^{(m)})=\textbf{E}[\ln
\textbf{f}(X^{(m)}|Y^{(m)})]\nonumber\\&& \DI(E_x\Rightarrow E_y) =
\sum_{m=1}^{M}\left(H(X^{(m)}|Y^{(m-1)})\right)
-\sum_{m=1}^M\left(H(X^{(m)}|Y^{(m)})]\right), \label{logit}
\end{eqnarray}
which gives the intuition that the DI is the cumulative reduction in
uncertainty of time sample $Y_m$ when the past time samples
$Y^{(m-1)}$ of $E_y$ are supplemented by information about the past
and present segments $X^{(m)}$ of $E_x$. In the case that the
feature sequences $X^{(M)}$ and $Y^{(M)}$ are jointly Gaussian, it
is easily shown that the DI reduces to a monotonic function of
Granger's linear causality measure.
\subsection{Previous SODA approach}
SODA quantizes the feature variables to $p$ levels denoted as
$\{z_1,\ldots, z_{p^m}\}$. If the feature realizations are i.i.d.
then $Z$ is multinomial distributed with probability mass function
\begin{eqnarray*}
&&P_\theta(Z_1=n_1,\ldots,Z_{p^m}=n_{p^m})=\frac{n!}{\prod_{k=1}^{p^m}n_{k}!}\prod_{k=1}^{p^m}\theta_{k}^{n_k},
\end{eqnarray*}
where $\theta=E[Z]/n=[\theta_1,\ldots,\theta_{p^m}]$ is a vector of
class probabilities and $\sum_{k=1}^{p^m}n_k=n$, where $n$ is
defined as the number of experimental trials,
$\sum_{k=1}^{p^m}\theta_k=1$. Since the number of quantization cells
$p^m$ is larger than the number of trials $n$, a brute force plug-in
estimation approach, e.g., using maximum likelihood (ML) estimates
in place of $\theta$, is prone to overfitting error. Specifically,
given $n$ independent samples $\{W_i\}_{i=1}^n$ of the EEG feature
vector $W=[X^{(M_x)},Y^{(M_y)}]$ the ML estimator of the $k$-th
class probability $\theta_k$ is $\hat{\theta}_k=n^{-1} \sum_{i=1}^n
I(W_i \in C_k)$, $k=1, \ldots, p^{M_x+M_y}$ where $I$ here is the
indicator function. In [37], SODA shrinkage regularization was
applied to reduce overfitting error.
\subsection{L-SODA extension}
To address the problem of high dimensionality of quantized feature
space, in this work we use multinomial logistic regression for
approximation of the conditional distribution [41] which obviates
the need to perform joint discretization of $X$ and $Y$. In
multinomial logistic regression, the logits,
\begin{eqnarray}
\log\frac{P(X=k|y)}{1+\sum_{x\neq
k}P(X=x|y)}=y^{T}\beta_k~,\label{app}
\end{eqnarray}
where $x$ and $y$ are discrete and continuous variables and
$\beta_k$ is determined by a goodness of fit criterion [41], are
modeled as a linear function. Using multinomial logistic regression
approximation (\ref{app}), the conditional probabilities required
for computation of (\ref{logit}) become,
\begin{eqnarray}
P(X^{(m)}=k|Y^{(m-1)})=\frac{\exp([Y^{(m-1)}]^{T}\beta_{k}(m-1))}{\sum_{j=1}^{p^{(m-1)}}\exp([Y^{(m-1)}]^{T}\beta_{j}(m-1))}~,
P(X^{(m)}=k|Y^{(m)})=\frac{\exp([Y^{(m)}]^{T}\beta_{k}(m))}{\sum_{j=1}^{p^m}\exp([Y^{(m)}]^{T}\beta_{j}(m))}~.
\end{eqnarray}
As compared to SODA, where both of $X^{(m)}$ and $Y^{(m)}$ are
discrete and there are ($|\mathcal C|^{M_x+M_y}$) of multinomial
parameters $\theta_k$. Here, in the L-SODA logistic regression
approach only $X^{(m)}$ is quantized. This results in a reduction of
multinomial dimensions from $|\mathcal C|^{M_x+M_y}$ to $|\mathcal
C|^{M_x}$, which is significant in EEG applications considered in
this paper. The regression coefficients $\beta=[\beta_1, \ldots,
\beta_{p^m}]$ are determined by maximum likelihood, with
$\beta_k(m)$ denoting the weight vector corresponding to class $k$
for $P(X^{(m)}|Y^{(m)})$. Thus, the estimated directed information
with multinomial logistic regression is:
\begin{eqnarray}
&&\widehat{DI}_{\beta}=\sum_{m=1}^{M}\left[\frac{\exp([Y^{(m-1)}]^{T}\hat{\beta}_{k}(m-1))}{\sum_{j=1}^{p^{m-1}}\exp([Y^{(m-1)}]^{T}\hat{\beta}_{j}(m-1))}\log
\frac{\exp([Y^{(m-1)}]^{T}\hat{\beta}_{k}(m-1))}{\sum_{j=1}^{p^{m-1}}\exp([Y^{(m-1)}]^{T}\hat{\beta}_{j}(m-1))}\right]\nonumber\\&&-\left[\frac{\exp([Y^{(m)}]^{T}\hat{\beta}_{k}(m))}{\sum_{j=1}^{p^{m}}\exp([Y^{(m)}]^{T}\hat{\beta}_{j}(m))}
\log
\frac{\exp([Y^{(m)}]^{T}\hat{\beta}_{k}(m))}{\sum_{j=1}^{p^{m}}\exp([Y^{(m)}]^{T}\hat{\beta}_{j}(m))}\right]
~,
\end{eqnarray}
Since $n \ll p^{\max\{|\mathcal C|^{M_x}\}}$, we propose to use
James-Stein shrinkage regularization to reduce the MSE of the DI
estimator. The resultant shrinkage optimized DI estimator,
$\widehat{DI}^{\lambda_{\circ}}(X^{M}\Longrightarrow Y^{M})$ shrinks
the maximum likelihood estimator $\hat{\beta}_{k}^{ML}$ towards a
target coefficient vector.
\begin{eqnarray}
\hat{\beta}_{k}^{\lambda}=\lambda
t_{k}+(1-\lambda)\hat{\beta}_{k}^{ML}~, \label{JS}
\end{eqnarray}
It is customary in James-Stein shrinkage to select targets that are
minimally informative, e.g. uniform density [5][37].
Correspondingly, we select $t_k$ so that
$t_{k,l}=\sqrt{\frac{1}{Var(y_l)}}$, namely, $t_{k,l}$ is inversely
proportional to the standard deviation of $y_l$, where $k$ is the
index of class and $t_{k,l}$ represents the $l$th element in the
target coefficient vector $t_k$. The James-Stein plug-in estimator
for directed information with logistic regression is:
\begin{eqnarray}
&&\widehat{DI}^{\lambda}_{\beta}=\sum_{m=1}^{M}[\frac{\exp([Y^{(m-1)}]^{T}\hat{\beta}_{k}^{\lambda}(m-1))}{\sum_{j=1}^{p^{m-1}}\exp([Y^{(m-1)}]^{T}\hat{\beta}_{j}^{\lambda}(m-1))}\log
\frac{\exp([Y^{(m-1)}]^{T}\hat{\beta}_{k}^{\lambda}(m-1))}{\sum_{j=1}^{p^{m-1}}\exp([Y^{(m-1)}]^{T}\hat{\beta}_{j}^{\lambda}(m-1))}\nonumber\\&&-\frac{\exp([Y^{(m)}]^{T}\hat{\beta}_{k}^{\lambda}(m))}{\sum_{k=1}^{p^{m}}\exp([Y^{(m)}]^{T}\hat{\beta}_{k}^{\lambda}(m))}
\log
\frac{\exp([Y^{(m)}]^{T}\hat{\beta}_{k}^{\lambda}(m))}{\sum_{j=1}^{p^{m}}\exp([Y^{(m)}]^{T}\hat{\beta}_{j}^{\lambda}(m))}]
~,
\end{eqnarray}
The corresponding plug-in estimator for DI is simply
$\widehat{\DI}^{\lambda}_{\beta}=\DI_{\hat{\beta}^\lambda}(E_x\rightarrow
E_y)$. We specify the optimal value of $\lambda$ that minimizes
estimator MSE: $\lambda^{\circ}=\arg\min_{\lambda}
E(\widehat{DI}^{\lambda}_{\beta}-DI)^{2}~\label{cost}$. The MSE can
be decomposed into the square of bias and variance and the optimal
value of $\lambda_{\beta}$ can be obtained by minimizing MSE over
$\lambda_{\beta}$ using a gradient descent algorithm.
\section{L-SODA-based Interaction Detection Algorithm}
In [37], a local version of DI was introduced for temporal
localization of interactions. This is an important step for studying
physiological signals such as the EEG, due to issues related to time
warping inherent in these data. Here we describe the L-SODA
algorithm in the context of EEG. Once the DI optimal shrinkage
parameter has been determined, the local DI is defined similarly to
the DI except that, for a pair of EEG signals $X$ and $Y$, the
signals are time shifted and windowed prior to DI computation.
Specifically, let $\tau_x\in [0,M_x-T]$, $ \tau_y\in [0,M_y-T]$ be
the respective time shift parameters, where $T \ll \min\{M_x ,M_y\}$
is the sliding window width, and denote  by $X_{\tau_x}^{M_x}$,
$Y_{\tau_y}^{M_y}$ the time shifted sequences. Then the local DI,
$\mathrm{DI}(X^{M_x}_{\tau_x} \rightarrow Y^{M_y}_{\tau_y})$,
computed using (\ref{DIformulationp}), defines a surface over
$\tau_x,\tau_y$. We use the peaks of the local DI surface to detect
and localize the interactions in the pair of EEG signals. We
implement L-SODA for EEG interaction detection using the following
procedure:
\begin{enumerate}
\item \textbf{Temporal Alignment:}  Align the EEG signals temporally by
segmenting the EEG signals according to local DI peak locations to
capture the beginning and ending times.

\item \textbf{Pairwise DI and
p-value computation:} After alignment, calculate the $K$ $\times$
$K$ matrices of SODA estimated DI's and  p-values $1-
\Phi\left(\frac{\hat{D}_{ij}-\mu_{ij}}{\sigma_{ij}}\right)$ on these
DI estimates here $K$ equals 19, the number of electrodes.
\item \textbf{False Discovery Rate Control:} Threshold
the DI and $p$-value matrices to find interaction regions exhibiting
large and statistically significant DI. The bootstrap is used to
estimate the mean and variance in the p-value matrix. The
construction of the interaction graph over the $K$ EEG electrodes is
performed by testing the $K \times (K-1)$ hypotheses that there is a
significant interaction (both directions) between pairs of
electrodes. Since there are $K(K-1)$ different DI pairs, this is a
multiple hypothesis testing problem and we control false discovery
rates using the corrected Benjamini- Hochberg (BH) procedure [35].
 It tolerates more
false positives, and allows fewer false negatives. The corrected BH
procedure is implemented as follows:
\begin{itemize}
\item The p-values of the $M=K(K-1)$ edges $(1,2,\ldots,M)$ are ranked from lowest to highest,
all satisfying the original significance cut-off $p=0.05$. The
ranked p-values are designated as $p_{(1)}$, $p_{(2)}$, $\ldots$ ,
$p_{(M)}$.
\item For $j = 1, 2, \ldots ,M$, the null hypothesis (no
edge) $H_j$ is rejected at level $\alpha$ if $p_{(j)}\leq
(j/m)/\left(\sum_{n=1}^M n^{-1}\right)\alpha$, where $\alpha$ is the
chosen acceptable p-value.
\item All the edges with p-value$\leq p_{(j)}$ are retained
in the final network.
\end{itemize}
\end{enumerate}
\section{Experimental Results}
\subsection{BCI Project Database} The L-SODA algorithm was applied to the public
BCI dataset consisting of EEG signals associated with motor activity
[17]. The EEG consists of random movements of the left and right
hand recorded with eyes closed. The data consisted of multiple data
matrices corresponding to multiple activities, where each column of
a data matrix represented one electrode and there are a total of 19
electrodes. Each row of a data matrix presented the temporal sample
of electrical potential from one electrode and there were a total of
3008 samples in each row. The motor activity lasted about 6 seconds.
The sampling rate of the recording was 500Hz. The data in BCI
project consisted of EEG signals. The subject executed 10 classes of
movements where each class contained different trials of the same
movement including three trials left hand forward movement, three
trials of left hand backward movement, three trials of right hand
forward movement, three trials of right hand backward movement, 1
trial of imagined left hand forward movement, 1 trial of imagined
right hand forward movement, 1 trial of imagined left hand backward
movement, 1 trial of imagined right hand backward movement, 1 trial
of left leg movement and 1 trial of right leg movement. The
application of L-SODA on these data was not trivial due to the time
misalignment, artifacts and noise variations. The time sequences
were first divided into segments of 200ms length for feature
extraction. There was 100ms overlap between neighboring segments. We
estimated the joint probability density functions for each segment
of EEG signal by first mapping the features to the codebook by
quantization. Then we applied the proposed shrinkage method to the
maximum likelihood estimator using. Here the number of samples $n$
was the total number of trials for all the subjects performing the
same task, and the codebook was learned in the training phase using
Lloyd-max quantization where the number of quantization levels in
the scalar quantizer was selected to be 10.
\subsubsection{Competing  algorithms investigated} The performance of
L-SODA-based interaction detection was compared to four state-of-the
art approaches: Granger's measure [2][27][31], coherency measure
[1], MI [3][4][21] and unregularized DI of Quinn et al.[26]. In [1],
coherency is defined as normalized cross-spectrum between two EEG
signals, where only the imaginary part of the signal was employed.
In [4][21], mutual information was applied to feature selection for
EEG signal classification. In [26], Quinn et al. utilized
unregularized directed information to capture the non-linear and
non-Guassian dependency structure of spike train recordings, where
they estimated the DI with point process theory and impose the
assumptions that the random processes are stationary and ergodic.
The DI estimator proposed by Quinn et al. was estimated as described
in [26]:
\begin{enumerate}
\item Find the parameters in generalized linear models (GLM) for
point processes [40] according to minimum description length (MDL)
procedure.
\item Calculate $\hat{H}(Y\parallel X)$ using generalized linear
models.
\item Compute an estimator for unconditional entropy rate $\hat{H}(Y)$ using a well-established entropy estimator
(such as Lempel-Ziv's estimator [39] or the Burroughs-Wheeler
Transform (BWT) based estimator [38]).
\item Calculate the directed information rate $\widehat{DI}(X\rightarrow Y)=\hat{H}(Y)-\hat{H}(Y\parallel
X)$.
\end{enumerate}
We implemented the generalized linear models for point process
relying on the code which is available at
$http://pillowlab.cps.utexas.edu/code\_GLM.html$. The classification
performance of L-SODA with kNN classifier was compared to the other
state-of-the-art approaches using MI [4] and Hidden Markov Models
[6], Granger's measure [23] and coherence measure [1]. The authors
of [14] proposed to address the signal classification problem by
combining hidden Markov model and maximum margin principle in a
unified kernel based framework called kernel based hidden Markov
model (KHMM).  In addition to the KHMM implementation, we
implemented the HMM by estimating the emission probability of the
distribution of EEG signals with Gaussian mixture models (GMM).
Specifically, for the GMM given 300 training trials and 300 test
trials, we implement the Baum-Welch algorithm with 50 iterations to
estimate the parameters of the GMM model governing EEG signals in
each activity class. A test EEG signal is classified using maximum
likelihood detected implemented by Viterbi's algorithm. For the
classification tasks, ground truths correspond to the labels for 10
different types of activities.  All data were divided into 2 sets of
50\% training and
50\% test samples each. \\
\subsubsection{Interaction Detection and Comparison}
\begin{figure}[h]
\begin{center}
\includegraphics[width=12.0cm,height=6.0cm]{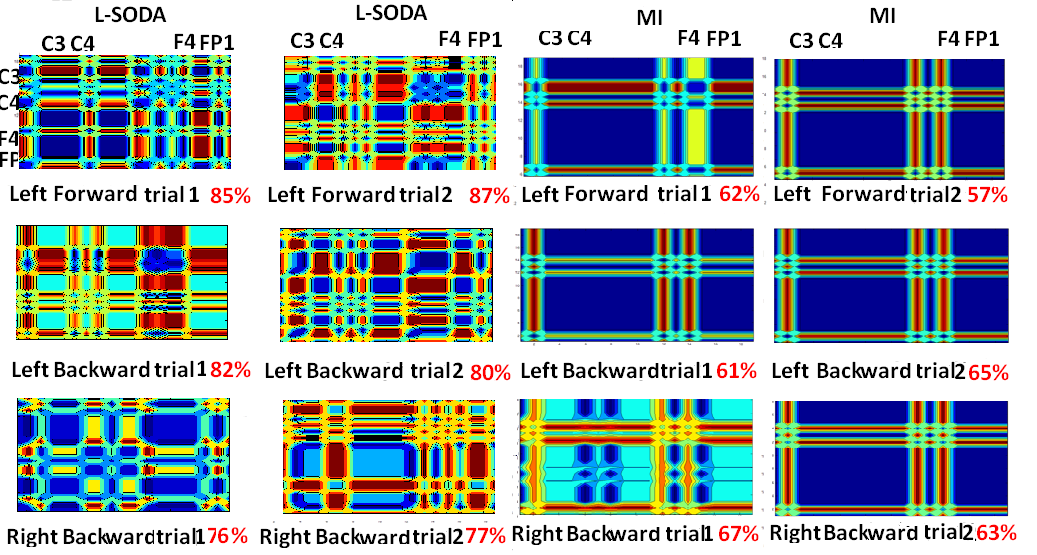}
\end{center}
\caption{Visual illustration of L-SODA and MI heatmaps with Project
BCI dataset for left hand forward movement, left hand backward
movement and right hand forward movement, where colors indicate the
magnitudes for different strengths of interactions calculated by
L-SODA and MI respectively between 19 electrodes in human brain.
L-SODA is more informative than MI for localization because L-SODA
is able to detect more interactions than MI, such as the
interactions between $FP1$ and $F_7$, $FP1$ and $FZ$. The red number
indicates the localization accuracy is computed by mapping the
detection results using SODA on different replicates of the same
class of activity. The order of the electrodes by clustering are
$T_3$, $T_5$, $C_3$, $C_4$,  $T_4$, $F_8$, $CZ$, $F_7$, $F_3$,
$T_6$, $PZ$, $P_3$, $P_4$, $O_1$, $O_2$, $FZ$, $F_4$, $FP_1$, $FP_2$
from left to right and from top to bottom.}\label{DIcontour}
\end{figure}

\textbf{Interaction Detection:} We evaluated the localization
performance using L-SODA. Fig.~\ref{DIcontour} was a visual
illustration of the DI matrix, expressed as a heatmap for left hand
forward movement, left hand backward movement and right hand forward
movement, respectively, where colors indicated the magnitudes for
different strengths of interactions between the 19 electrodes. In
the interest of space, we discuss the results for left hand forward
movement in more detail below (these results are representative of
other movements).

We used a heat kernel to transform (symmetrized) pairwise DI matrix
into the distance matrix and applied K-means clustering to the
distance matrix with the number of cluster (3) chosen by setting a
threshold to within-cluster sums of point-to-centroid distances. For
left hand forward movement, the three clusters were $(C_3, C_4, T_3,
T_5)$, $(F_4, FP1, FP2)$ and the rest of the electrodes. Mapping the
EEG channels into Brodmann areas [25], we identified cluster $(C_3,
C_4, T_3, T_5)$ as reflecting auditory processing such as that
associated with detecting a cue to start motion (Brodmann area 21)
and the execution of motor function (Brodmann area 4). Similarly, we
identified cluster $(F_4, FP1, FP2)$ as corresponding to the
planning of complex movements (Brodmann area 8) and cognitive
branching (Brodmann area 10). The third cluster corresponded to
electrodes that were not very active. The localization accuracy was
highlighted for each heatmap with red numbers in
Fig.~\ref{DIcontour}. In Fig.~\ref{DIcontour}, the order of the
electrodes by clustering were $T_3$, $T_5$, $C_3$, $C_4$, $T_4$,
$F_8$, $CZ$, $F_7$, $F_3$, $T_6$, $PZ$, $P_3$, $P_4$, $O_1$, $O_2$,
$FZ$, $F_4$, $FP_1$, $FP_2$ from left to right and from top to
bottom. Compared to MI, L-SODA was more informative for localization
with 10\% improvement on average and is able to detect more accurate
interactions when the same threshold of p-values was applied, where
the localization accuracy has been highlighted along with the
heatmap using red numbers. While MI and coherence measure [1][4]
were not able to detect such interactions due to lower sensitivity
to directional information flow.
\begin{figure}[h]
\begin{center}
\includegraphics[width=8.5cm,height=8.5cm]{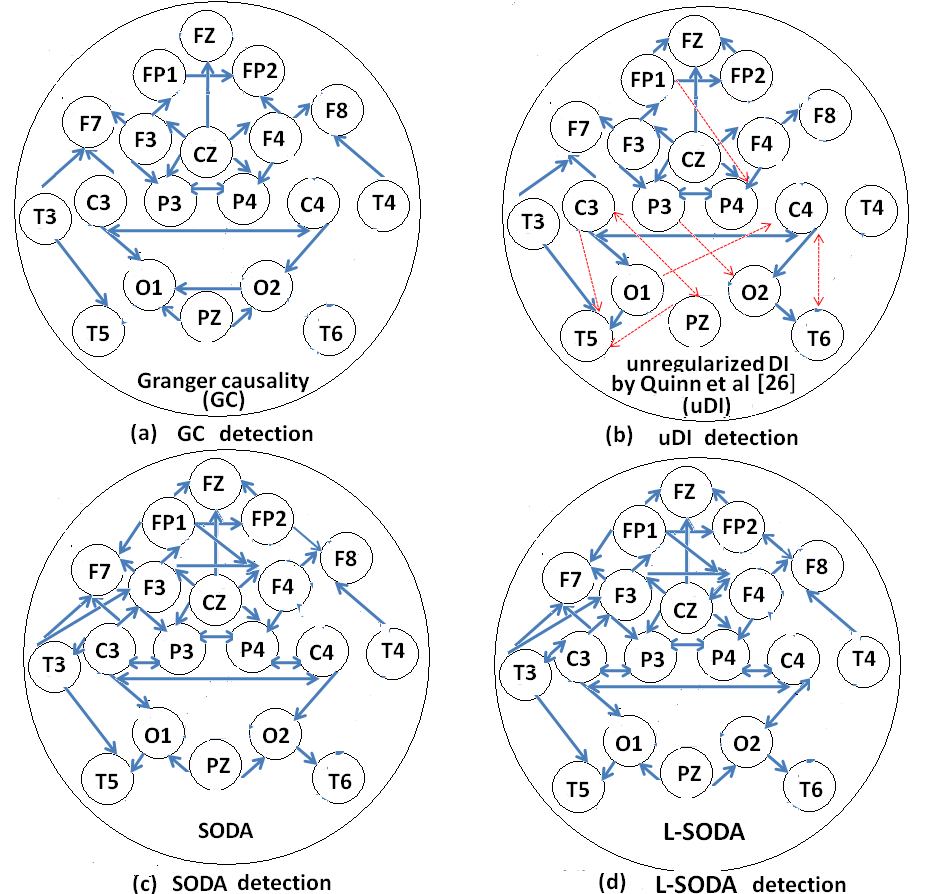}
\end{center}
\caption{Visual illustration of the dependencies between different
electrodes for the activity "left hand forward movement"
reconstructed using Granger's measure, unregularized DI by Quinn et
al.[26], SODA and L-SODA averaged over the total 6 seconds, where
the threshold for declaring an edge present corresponds to a p-value
of level 0.05 which corresponds to the best localization accuracy
shown in Table 1. The False Discovery Rate (FDR) is controlled below
0.1 using corrected BH procedure to resolve the multiple comparison
problem and the sliding window width is $T=7$. Again, L-SODA detects
more interactions compared to MI, Granger's measure and
unregularized DI. For instance, the edges between $(FP_1 \rightarrow
F_4)$ and $(FP1 \rightarrow FZ)$ corresponding to p-values of level
0.027 and 0.035 respectively using L-SODA. These interactions cannot
be discovered by other approaches. The red dot arrows are false
positives with unregularized DI by Quinn et al.[26] validated by
neural pathway locations and Brodmann area, where the channels
denoted as false positives do not have motor-related
functions.}\label{DInetwork}
\end{figure}

\textbf{Comparison:} Fig.~\ref{DInetwork} compared interaction graph
discovery obtained through SODA with other state-of-the-art
approaches using the activity "Left Hand Forward Movement". These
results were representative but, due to space limitations, the
interaction graphs for other motor activities are not discussed
here. The regions in the brain for Fig.~\ref{DIcontour} that exhibit
the highest activity match perfectly with the regions in
Fig.~\ref{DInetwork} that have densest number of links.
Fig.~\ref{DInetwork} indicated that L-SODA discovered significantly
more interactions where the sliding window size is $T=7$. These new
interactions can be validated by the fact that p-values are
significant. For instance, the edges between $(FP1 \rightarrow F_4)$
and $(FP1 \rightarrow F_7)$ corresponded to p-values of level 0.027
and 0.035 respectively. The results of applying SODA on the
replicates of EEG signals also indicated that during these periods,
the electrodes $FP1$, $F_4$ and $F_7$ were highly interactive and
therefore can serve as strong evidence that the activity was indeed
being localized to these electrodes in the brain associated with
motor control. Compared to the unregularized DI [26], L-SODA has the
advantage that it can control false positive rate more accurately
and its predictions are validated by neural pathway locations as
determined by Brodmann areas. The main reason that L-SODA has the
superior performance compared to the unregularized DI [26] is
because the unregularized DI may underestimate the DI in the
presence of small number of samples and high dimensional signals.

\begin{figure}[h]
\begin{center}
\includegraphics[width=10.0cm,height=7.0cm]{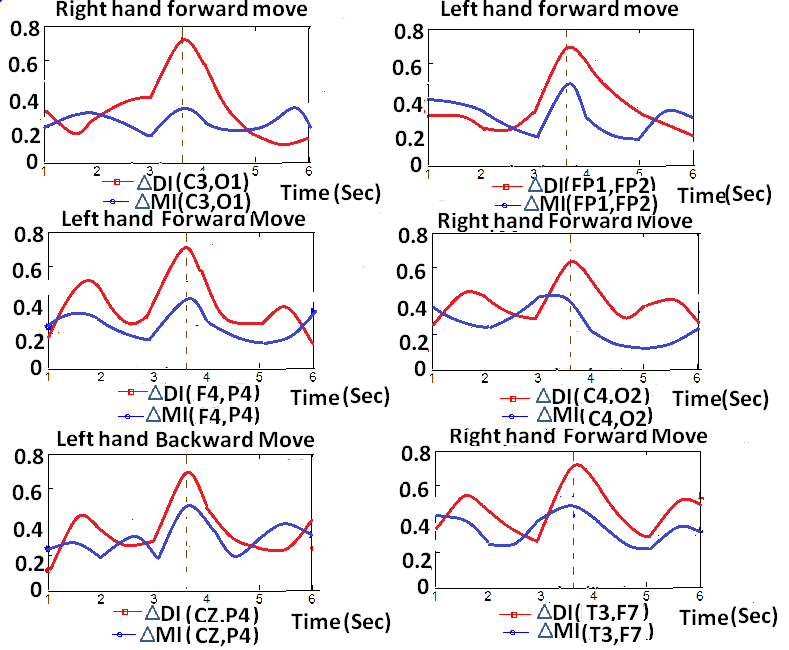}
\end{center}
\caption{Illustration of DI as compared to MI for capturing
similarities between activities for six pairs of EEG signals from
different sources of electrodes in BCI project dataset such as
$(C_3,O_1)$, $(FP_1,FP_2)$. As it is a directional measure, DI is
more sensitive than MI to the emergence of the mental task and this
can be seen from the fact that the peak of red $\Delta$DI trajectory
is sharper than the peak of the blue $\Delta$MI trajectory over
time. ($\Delta$DI is the temporal change of the DI similarity
measure over successive temporal segments and similarly for
$\Delta$MI). The dashed vertical lines correspond to the beginnings
of the activities. }\label{NIPSDI}
\end{figure}

\begin{figure}[h]
\begin{center}
\includegraphics[width=12.0cm,height=6.0cm]{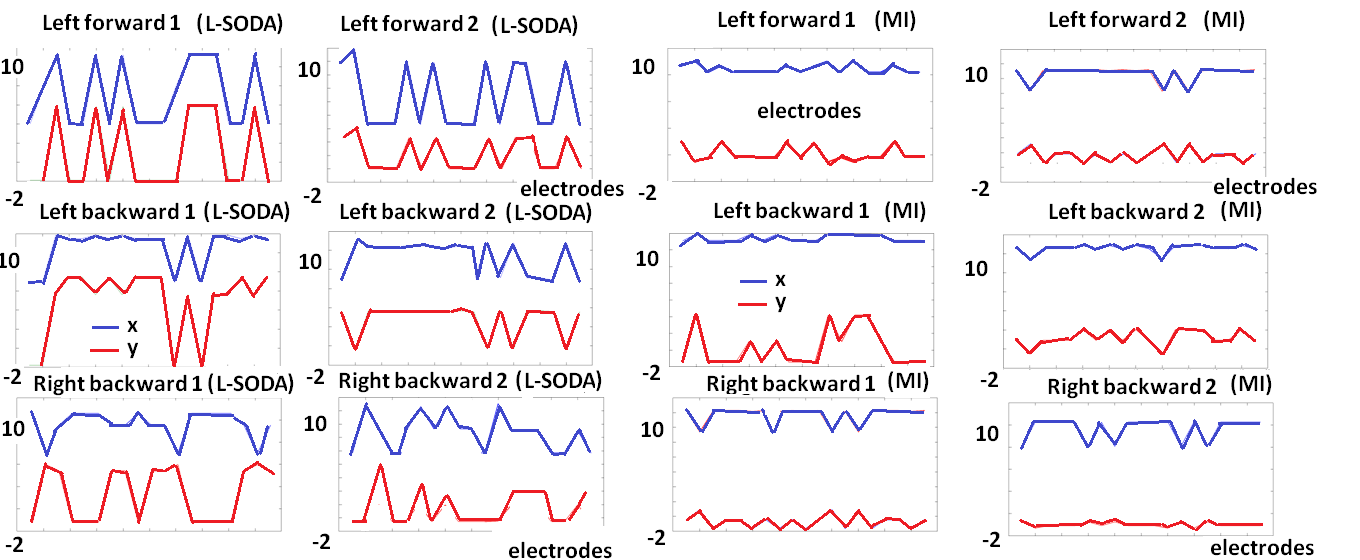}
\end{center}
\caption{Visual illustration of electrodes by applying
multidimensional scaling on the heatmaps and reducing the dimensions
into 2 with L-SODA and MI for different activities, where the blue
and red curves represent the first dimension and the second
dimension. The horizontal axis in the figure represents the
electrodes and the vertical axis represents the amplitudes. By
comparing these dependency scatterplots to the spatial organization
of the electrodes, interesting patterns are identified for different
activities. For instance, the electrodes $C_3$ and $C_4$, $T_3$ and
$T_5$ have strong interactions for left hand forward movement. Some
of the representative traces are shown in
Fig.~\ref{NIPSDI}.}\label{mds}
\end{figure}

In Fig.~\ref{NIPSDI} we plot the local DI and local MI as time
trajectories. These trajectories can be interpreted as scan
statistics for localizing interactions in the two EEG signals.
Fig.~\ref{NIPSDI} illustrated of the advantages of DI as compared to
MI for capturing similarities between six pairs of EEG signals from
BCI dataset. As shown in Fig.~\ref{NIPSDI}, DI is more sensitive
than MI to the emergence of the mental task due to the fact that DI
is a directional measure. The results show that DI can successfully
detect highly interactive periods between pairs of electrodes
corresponding to the temporal annotations in BCI dataset. Similar
performance can be shown for the rest pairs of interactions in the
total 19$\times$ 18=342 pairs of possible interactions.\\
\textbf{Spatial Pattern:} To further reveal the spatial patterns of
the electrodes, in Fig.~\ref{mds}, we show visual illustration of 2D
scatter plots of electrodes by applying multidimensional scaling on
the heatmaps and reducing the dimensions into 2 with L-SODA and MI
for different activities, where the blue curve represented the first
dimension and the red curve represents the second dimension. By
comparing these dependency scatterplots to the spatial organization
of the electrodes, interesting patterns were identified for
different activities. For instance, the electrodes $C_3$ and $C_4$,
$T_3$ and $T_5$ had strong interactions for left hand forward
movement.
\begin{figure}[h]
\begin{center}
\includegraphics[width=14.0cm,height=10.0cm]{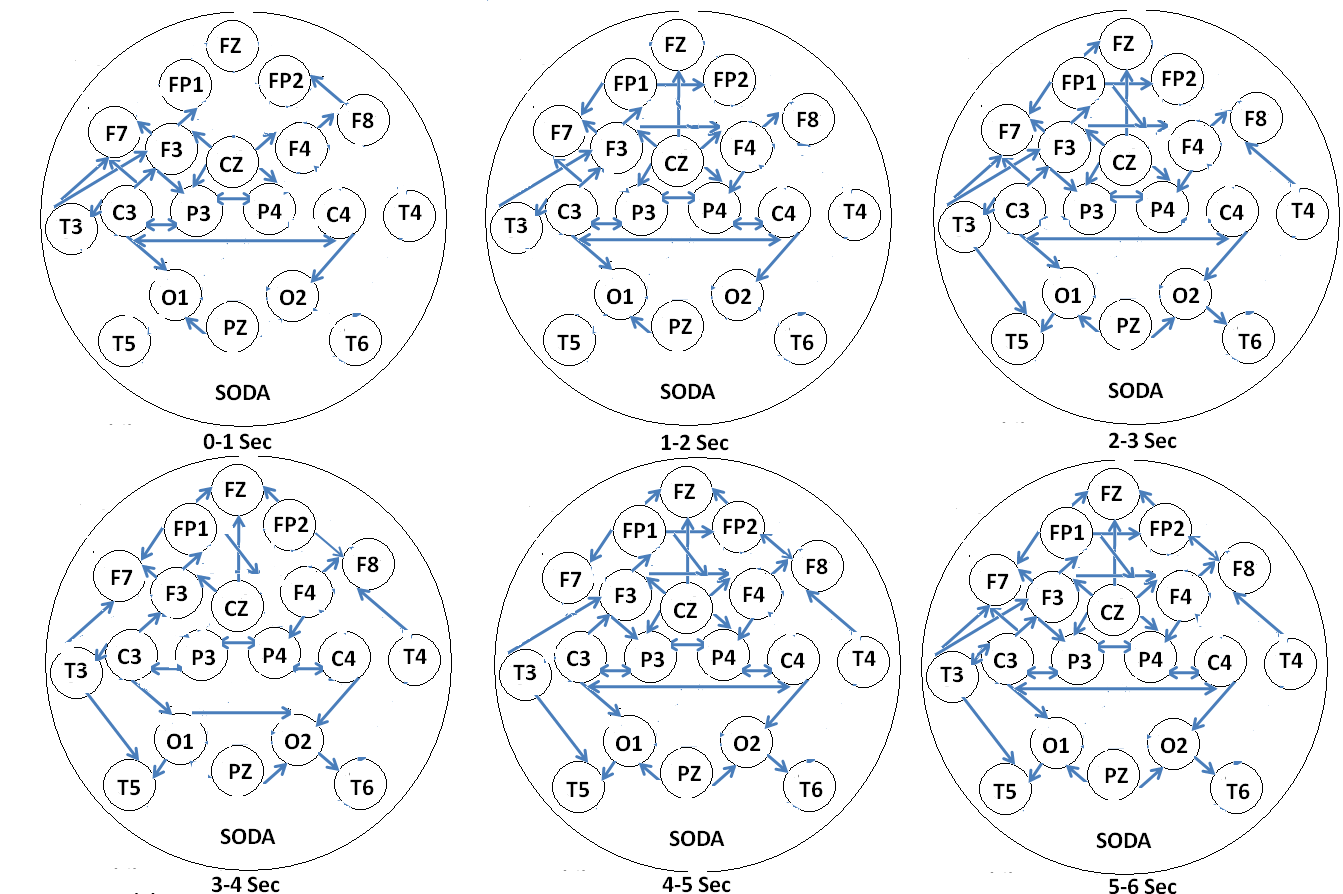}
\end{center}
\caption{Visual illustration of the dependencies between different
electrodes for the activity "left hand forward movement"
reconstructed using L-SODA dynamically with 1 second interval, where
the results demonstrate that the interactions first start from the
regions close to C3 and C4 which is Somatosensory and Motor region,
and then transmit to other regions in the human
brain.}\label{DInetwork2}
\end{figure}

\textbf{Dynamical Analysis:} Besides the average localization and
interaction detection performance, it was more desirable to reveal
the dynamical process of the interaction between the EEG channels.
To demonstrate the interactions among electrodes dynamically,
Fig.~\ref{DInetwork2} plots visual illustration of the dependencies
between different electrodes for the activity "left hand forward
movement" reconstructed using L-SODA with 1 second interval, where
the results demonstrated that the interactions first start from the
regions close to C3 and C4 which is Somatosensory and Motor region,
and then transmit to other regions in the human brain.

\subsubsection{Consistency Measure}
To study the ability of SODA to uncover interactions that were
consistently observed during the same class of activity, we randomly
divided the data into equal sized training and test sets. SODA was
applied to the training set and the localization consistency was
computed by mapping the detected interactions to the testing set.
Table 1 demonstrated the comparison of the EEG localization accuracy
of different algorithms using different levels of false discovery
rates (FDR) including MI, Granger's measure where the covariance
matrix is regularized with Ledoit Wolf shrinkage method, coherence
measure, unregularized DI of Quinn et al. and L-SODA. A localization
consistency of 100\% means that all interactions discovered on the
training set were observed in the test set. As shown in Table 1,
when the level of FDR was large, the five methods have comparable
performance. When the threshold was lowered, L-SODA significantly
outperforms other methods in localization accuracy. Since the weak
dependencies were filtered out and strong dependencies remains with
a lower threshold, false positives are reduced for all the five
methods. Meanwhile, the hit using L-SODA remained to be high due to
its non-symmetric measure. This explains the fundamental reason that
L-SODA achieved the best performance with the threshold for
declaring an edge present corresponding to a p-value of level 0.05.
Compared to Granger's measure and unregularized DI which were
non-symmetric measures, L-SODA improved the accuracy by about 8\%
and 3\% due to the fact that L-SODA was a shrinkage optimized
version of directed information estimator for non-Gaussian
distribution of the signals.

We also assessed localization consistency in terms of interactions
within regions of the brain coordinating similar task related
behavior (i.e., as opposed to between all EEG electrode sites).
Using the corrected BH procedure [35] for L-SODA, 39 interactions
were detected when FDR is equal to 0.1, while 24, 27 and 32
interactions were detected using MI, CM and unregularized DI. In
Table 2, we identified electrodes that are either sources,
recipients, or both sources and recipients of information flow from
the second-by-second plots of the 19$\times$19 matrices. Table 3
demonstrated the Brodmann areas and the corresponding functions for
the channels that are either sources, recipients, or both sources
and recipients of information flow, where we identified that the
directed information graph discovered by L-SODA is consistent with
activation of the known Brodmann areas of the brain associated with
motor functions. The comparison of the average running time for
directed information between pairs of channels is reported in Table
4, where we demonstrate that L-SODA has the same order of
computational complexity compared to the unregularized DI by Quinn
et al.[26].

\subsubsection{Classification} 
\begin{figure}[h]
\begin{center}
\includegraphics[width=12.0cm,height=7.0cm]{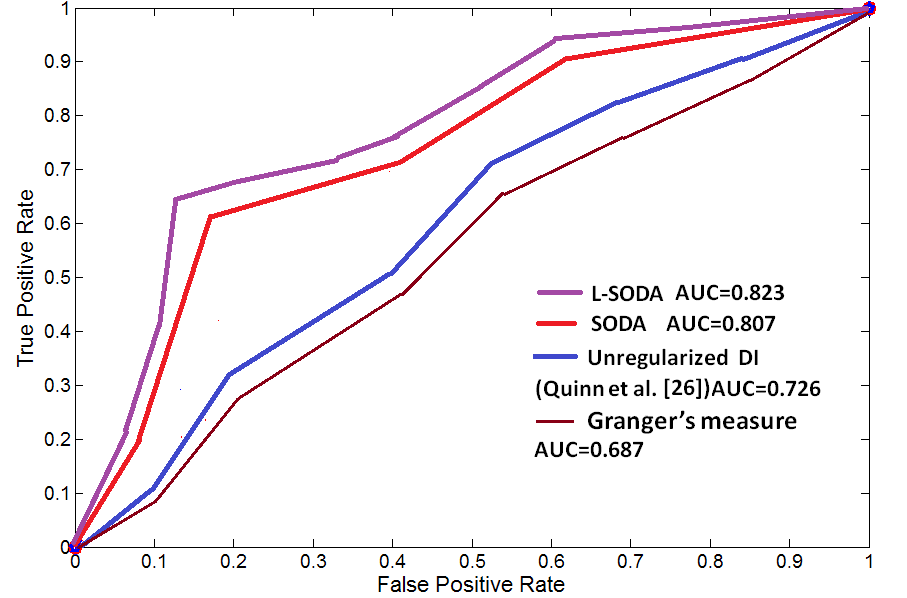}
\end{center}
\caption{Comparisons of ROC curves of classification performance for
BCI dataset with all the 10 classes of activities using L-SODA,
SODA, Granger's measure and unregularized DI by Quinn et al.[26],
where L-SODA significantly outperforms Granger's measure and the
unregularized DI by Quinn et al.[26] in terms of area of the curve
(AUC).}\label{ROC}
\end{figure}

We next evaluated the classification
performance of EEG signals using L-SODA as compared to MI and Hidden
Markov Model (HMM) where the objective was to classify among the 10
classes of activities from the EEG data. The L-SODA classification
task was conducted by applying the $k$ nearest neighbor classifier
on the pair-wise distances computed using L-SODA. In Table 5 we
compared the classification performance of L-SODA to that of the
unregularized DI, the HMM implemented with GMM and Kernel-based
Hidden Markov Model (HMM) where Gaussian radial basis function was
utilized evaluated using EEG signals in the BCI project dataset.
Table 5 indicated L-SODA outperforms HMM and unregularized DI in
terms of EEG signal classification. This improvement may be
attributed to the presence of model mismatch and bias in the HMM
model as contrasted to the more robust behavior of the proposed
model-free shrinkage DI approach. A more comprehensive quantitative
comparison was shown in Table 6 with the mean and standard
deviations for different activities. The superior performance of
L-SODA compared to coherence measure and MI can be attributed to the
fact that coherence measure and MI were symmetric measures and less
sensitive in capturing the directional information flow between EEG
signals such as the directed dependencies between ($T3\rightarrow
F7$), ($T4\rightarrow F8$). Granger's measure was based on a strong
Gaussian model assumption, which may account for its inferior
performance. In Fig.~\ref{ROC}, we demonstrated the comparisons of
ROC curves for classification performance with all the 10 classes of
activities using L-SODA, SODA, Granger's measure and unregularized
DI by Quinn et al.[26], where L-SODA significantly outperformed
Granger's measure and the unregularized DI by Quinn et al.[26] in
terms of area under the curve (AUC). L-SODA achieved the AUC 0.823,
Granger's measure achieved the AUC 0.687 and the unregularized DI
achieves the AUC 0.726. L-SODA has 8\% lower false negative rate
(namely 8\% higher in true positive rate) in detecting significant
information flow at given level of false positives in terms of ROC
curves, which can be mainly attributed to the fact that the use of
optimized shrinkage regularization estimator in L-SODA.

\begin{table}[!htb]
\begin{tabular}{|c|c|c|c|}
  \hline
  FDR & 0.1 & 0.07 & 0.05  \\
  \hline
  MI & 0.641 & 0.657 & 0.676 \\
  \hline
  GC & 0.653 & 0.705 & 0.728 \\
  \hline
  CM & 0.657 & 0.694 & 0.726 \\
  \hline
  uDI & 0.669 & 0.721 & 0.743 \\
  \hline
  SODA & \textbf{0.698} & \textbf{0.755} & \textbf{0.809}\\
  \hline
  L-SODA & \textbf{0.715} & \textbf{0.767} & \textbf{0.823}\\
  \hline
\end{tabular}
 \caption{Comparison of the EEG localization accuracy for different
 level
 of significance (false discovery rate), where the accuracy is computed by mapping the detection
results using L-SODA on different replicates and the number of these
electrodes that are connected in the MI, GC, CM, uDI, SODA and
L-SODA interaction graphs, determined by thresholding these
quantities at the same FDR level. uDI, CM and GC represents
unregularized DI, coherence measure and Granger
causality.}\label{HMM} \centering
\end{table}

\begin{table}
\begin{tabular}{|c|c|c|c|c|c|}
  \hline
   & 0-1s & 1-3s & 3-4s & 4-5s & 5-6s\\
  \hline
  Send & FP2 & FP2 & $T_4, F_8$ & Null & Null \\
  \hline
  Send and Receive & $F_8, T_4, T_5$ & $F_8, T_4$ & FP2 & FP1,FP2,CZ & $T_4,T_5,F_8,FP2,CZ$ \\
  \hline
  Receive & FZ & $P_3, T_6$ & $P_3, T_6$ & $F_3,F_4,T_3,T_4,T_5,F_8$  & $T_3$ \\
  \hline
\end{tabular}
\caption{Identification of the electrodes that are either sources,
recipients, or both sources and recipients of information flow from
the second-by-second plots in Fig.~\ref{DInetwork2}.}\label{dynamic}
\centering
\end{table}

\begin{table}
\begin{tabular}{|c|c|c|}
  \hline
 Channel  & Brodmann Area & Function \\
  \hline
  FP2 & right 10 & executive function; tertiary motor (E) \\
  \hline
  $F_8$ & right 44,45,46,47 & grasping/manipulation; tertiary motor (G) \\
  \hline
   $T_4$ & right 21  & contemplating distance (C)\\
  \hline
   $T_3$ & left 21  & contemplating distance (C) \\
  \hline
   $CZ$ & middle 5  & maintain spatial reference for goal oriented
   behavior (MA) \\
   \hline
   $FP1$ & left 10  & executive function; tertiary motor (E)
 \\
 \hline
  $T_5$ & left 19,37  & motion sensitive visual processing; contemplating
  distance (MO)
 \\
 \hline
  $F_3$ & left 8  & planning of complex movements (P)
 \\
 \hline
  $F_4$ & right 8  & planning of complex movements (P)
 \\
 \hline
  $T_6$ & right 19,37  & motion sensitive visual processing; contemplating
  distance (MO)
 \\
 \hline
\end{tabular}
\caption{The Brodmann areas and the corresponding functions for the
channels that are either sources, recipients, or both sources and
recipients of information flow from the second-by-second plots in
Fig.~\ref{DInetwork2}.}\label{dynamic} \centering
\end{table}

\begin{table}[!htb]
\begin{tabular}{|c|c|c|}
  \hline
   & unregularized DI by Quinn et al.[26] & L-SODA \\
  \hline
  CPU time (sec) & 0.12 & 0.15 \\
  \hline
\end{tabular}
\caption{Comparisons of the average running time for directed
information between pairs of channels using L-SODA and unregularized
DI by Quinn et al.[26] for BCI dataset, where L-SODA and
unregularized DI have the same order of computational
complexities.}\label{HMM} \centering
\end{table}

\begin{table}[!htb]
\begin{tabular}{|c|c|c|c|c|}
  \hline
   & HMM(n=2) & HMM(n=5) & HMM(n=7) & KHMM [14] \\
  \hline
  AP & 0.645 & 0.683 & 0.712 & 0.749  \\
  \hline
  & MI & unregularized DI & SODA & L-SODA \\
  \hline
 AP & 0.661 & 0.768 & \textbf{0.815} & \textbf{0.826} \\
  \hline
\end{tabular}
\caption{Comparisons of Average Precision (AP) for EEG signal
classification for logit shrinkage optimized directed information
assessment (L-SODA), SODA, unregularized DI, Hidden Markov Model
(HMM) with Gaussian mixture model (GMM) ($n$ is the number of
components) and kernel-based Hidden Markov Model (KHMM), where the
average precision is calculated by averaging the classification
accuracy over all the activities. Ground truths correspond to the
labels for 10 different types of activities.}\label{HMM} \centering
\end{table}

\begin{table*}[!htb]
\begin{tabular}{|c|c|c|c|c|}
  \hline
  Task & Left Hand Forward [\%] & Right Hand Forward [\%] & Left Leg [\%] & Right Leg [\%]\\
  \hline
  Granger's measure & 69.3$\pm$ 5.3 & 72.5$\pm$ 3.6 & 67.6$\pm$ 2.1 & 63.5$\pm$ 2.6 \\
  \hline
  Coherence Measure[1] & 73.5$\pm$ 2.7 & 74.1$\pm$ 4.2 & 68.1$\pm$ 8.3 & 65.8$\pm$ 5.1 \\
  \hline
  Mutual Information & 65.1$\pm$ 2.9 & 65.9$\pm$ 3.0 & 63.0$\pm$ 11.3 & 62.5$\pm$ 3.8 \\
  \hline
  Quinn et al's DI [26] & 72.1$\pm$ 4.0 & 75.6$\pm$ 2.3 & 70.3$\pm$ 8.5 & 67.9$\pm$ 6.9 \\
  \hline
  SODA & \textbf{81.3$\pm$ 4.6} & \textbf{81.7$\pm$ 2.5} & \textbf{76.1$\pm$5.8} & \textbf{77.5$\pm$ 4.3} \\
  \hline
  L-SODA & \textbf{82.5$\pm$ 5.1} & \textbf{83.1$\pm$ 4.5} & \textbf{78.2$\pm$ 4.7} & \textbf{79.5$\pm$ 3.3} \\
  \hline
\end{tabular}
\label{table:accuracy} \caption{Percentages (mean and standard
deviations) of correctly recognized single trials for the different
activities for Project BCI dataset. L-SODA gives at least 6\%
improvement in average precision over the best alternative Quinn et
al's regularized DI and L-SODA has roughly similar performance
compared to SODA. The $k$ nearest neighbor classifier is implemented
for classification.} \centering
\end{table*}

\subsection{CHB-MIT Scalp EEG Database}
The CHB-MIT Scalp EEG Database, collected at the Children's Hospital
Boston, consists of EEG recordings from pediatric subjects with
intractable seizures. Subjects were monitored for up to several days
following withdrawal of anti-seizure medication in order to
characterize their seizures and assess their candidacy for surgical
intervention. All signals were sampled at 256 samples per second
with 16-bit resolution. The recordings, grouped into 23 cases, were
collected from 22 subjects. Different from BCI dataset, the CHB-MIT
Scalp EEG Database has the annotations of the beginning and the
ending time of the onset of seizure. The main task associated with
the database is seizure detection. There are three performance
metrics of interest. The electrographic seizure onset detection
latency $EO_{latency}$ corresponds to the delay between
electrographic onset and detector recognition of seizure activity.
The sensitivity represents to the percentage of test seizure
identified by a detector. The false alarm per hour is the number of
times, over the course of an hour, that a detector declares the
onset of seizure activity in the absence of an actual seizure.
Generally, the goal of seizure detection is to signal an alert
within 10 seconds of seizure onset. The non-seizure vectors are
computed from at least 24 hours of nonseizure EEG, where both of
awake and sleep status for non-seizure EEG are included. In our
work, we compared the use of L-SODA to the best known seizure
detector based on Shoeb et al. [42]. The approach of Shoeb et al.
uses energy-based features obtained by passing the EEG signals
through $M$-band filterbank $(M=8)$ through 0.5-24Hz and measuring
the energy in the subband signals. In our experiments, detection was
formulated as a binary classification problem (seizure vs.
non-seizure) and implemented with support vector machine (SVM)
classification. In order to formulate the L-SODA feature vector, we
pre-processed the data by using a bandpass filter from 0.5 to 70 Hz,
with a notch filter at 59-61 Hz to remove line noise. Subsequently,
we applied multidimensional scaling (MDS) on the distance matrices
estimated by L-SODA to reduce them to 2 dimensions. The SVM used by
the detector is trained using the LibSVM software package with a
cost factor $J=1$, RBF kernel parameter of $\gamma=0.1$ and
trade-off between classification margin and error $C=1$ with 2-fold
cross-validation. \\

\begin{figure}[h]
\begin{center}
\includegraphics[width=12.0cm,height=7.0cm]{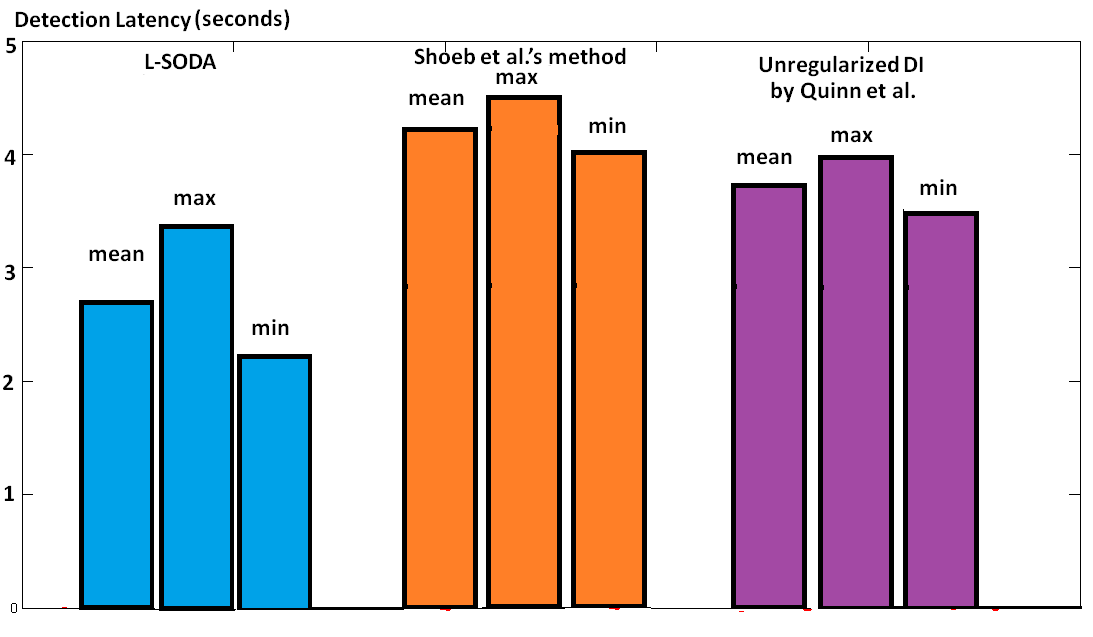}
\end{center}
\caption{Comparison of the mean, minimum and maximum of detection
latency (seconds) for pre-seizure data using L-SODA, the method by
Shoeb et al. [42] and unregularized DI by Quinn et al.
[26].}\label{latency}
\end{figure}

\begin{figure}[h]
\begin{center}
\includegraphics[width=12.0cm,height=7.0cm]{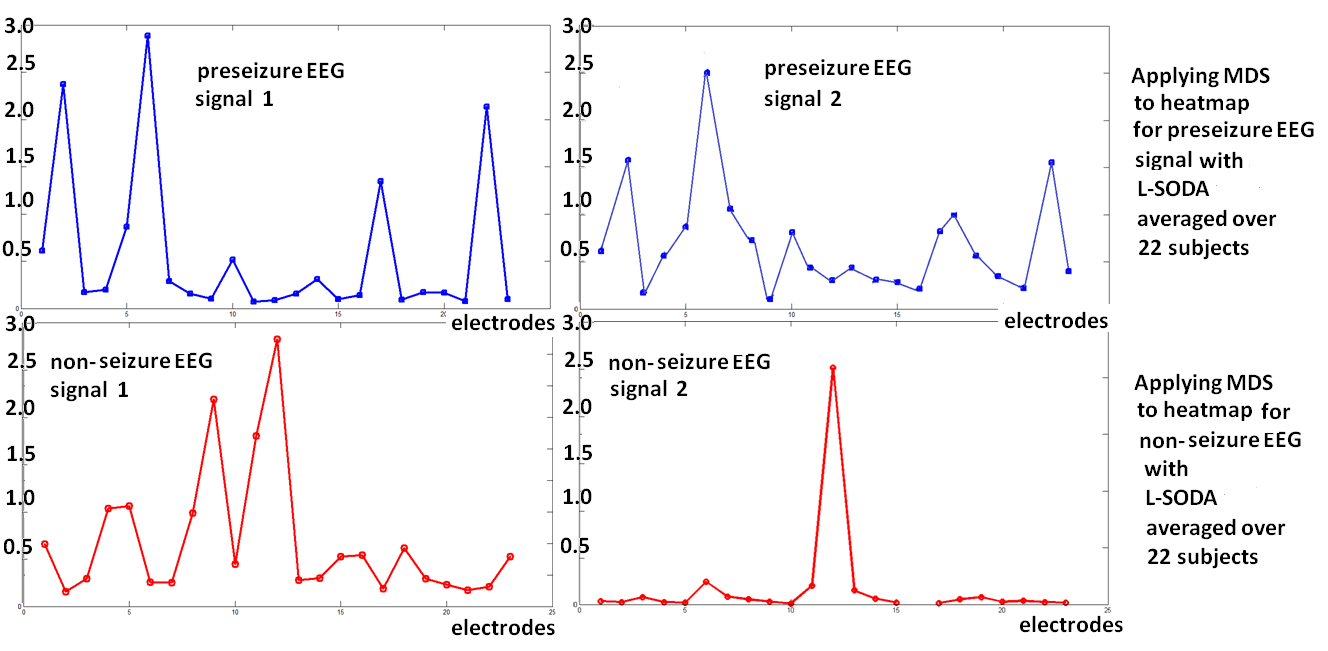}
\end{center}
\caption{Comparison of the results for applying multi-dimensional
scaling (MDS) to heatmaps for pre-seizure and non-seizure EEG
signals, where the result is averaged over 22 subjects and the
pre-seizure data represents the scalp EEG signal within 10 seconds
before the onset of seizure. The vertical axis represents the
amplitude and the horizontal axis represents the
electrodes.}\label{pattern1}
\end{figure}

\begin{figure}[h]
\begin{center}
\includegraphics[width=12.0cm,height=7.0cm]{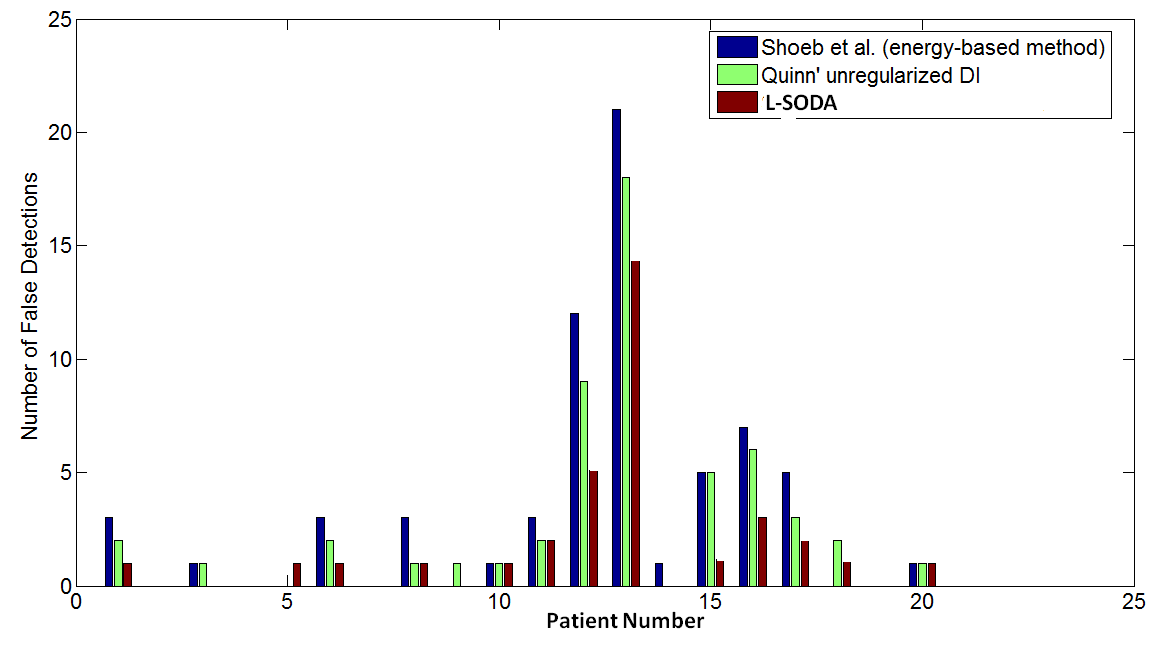}
\end{center}
\caption{Comparison of false detections for seizure data using the
patient-specific detector with L-SODA, unregularized DI by Quinn et
al. [26] and energy-based method by Shoeb et al. [42] where $x$ axis
represents the patient number and $y$ axis represents the number of
false detections.}\label{FAcompare}
\end{figure}

Overall, 97.1\% out of the 173 test seizures were detected using
L-SODA, which was slightly better than the energy-based method by
Shoeb et al. with 96\%. The mean latency with which the L-SODA
detector declared seizure onset was 2.8 seconds. We demonstrate in
Fig.~\ref{latency} the comparison of the mean, minimum and maximum
detection latency for pre-seizure data using L-SODA, the method by
Shoeb et al. [42] and unregularized DI by Quinn et al. [26], where
L-SODA had the shortest detection latency due to its sensitivity for
temporal changes of EEG signals. In Fig.~\ref{pattern1} we
demonstrated the comparison of the results for applying MDS to
heatmaps for pre-seizure and non-seizure EEG signals where the
result was averaged over 22 subjects and the pre-seizure data
represents the scalp EEG signal within 10 seconds before the onset
of seizure where the vertical axis represented the amplitude and the
horizontal axis represents the electrodes. As shown in
Fig.~\ref{pattern1}, the patterns identified by L-SODA for
pre-seizure and non-seizure EEG signals were significantly
different, which resulted in accurate prediction when SVM is
implemented. In Fig.~\ref{FAcompare}, we demonstrated that
comparison of false detections for the patient-specific detector
using L-SODA, unregularized DI by Quinn et al. and energy-based
method by Shoeb et al. where $x$ axis represented the patient number
and $y$ axis represents the number of false detections. While we
found that all of the three methods can achieve high positive rates,
as shown in Fig.~\ref{FAcompare}, the proposed L-SODA with logistic
regression have reduced the number of false detections compared to
unregularized DI by Quinn et al. [26] and Shoeb et al.'s energy
feature-based method especially for the patients 1, 6, 12, 13, 15,
16 and 17. The superior performance can be mainly attributed the
fact that the L-SODA is more sensitive in capturing the patterns of
directional information flow while the energy feature-based feature
described by Shoeb et al. does not account for the temporal
dependency of the EEG signals. Therefore, their method misses or has
a large detection latency when a test seizure differs significantly
in spatial or spectral character from all of the seizures in the
training set.

\section*{Conclusion}
We proposed a novel non-parametric model-free framework called
L-SODA for EEG signal interaction detection and classification based
on directed information. L-SODA uses a new James-Stein shrinkage
approach to logistic regression of directed information estimation
resulting in minimum mean squared error. A central limit theorem for
the L-SODA estimator specifies p-values that can be used to filter
out false positive peaks of the estimated L-SODA. We illustrated the
L-SODA estimator for EEG signals interaction detection/localization
and classification using BCI databases. L-SODA provides interaction
estimated that are consistent with neural pathway locations as
determined by Brodmann areas. Our results indicate that L-SODA is
able to detect interaction regions in human brains that involve
strong directional information flow without imposing strong model
assumptions. Since L-SODA captures the directional information that
EEG signals naturally possess and successfully controls the
overfitting error with optimized shrinkage regularization, L-SODA
demonstrates better performance as compared to unregularized DI and
undirected methods such as MI or coherence measure methods.
Moreover, we evaluated the L-SODA on CHB-MIT Scalp EEG database for
seizure detection. We demonstrated that compared to the
state-of-the-art approaches, the proposed method provides better
performance in detecting the epileptic seizure.

\section*{Acknowledgments}
This work was partially supported by US Army Research Office (ARO)
grant W911NF-09-1-0310 and NSF grant IIS-1054419. No additional
external funding received for this study.
\section*{Reference}
{[1] G.~Nolte, O.~Bai, L.~Wheaton, Z.~Mari, S.~Vorbach and
M.~Hallett, {\it Identifying true brain interaction from EEG data
using the imaginary part of coherency}, Clinical Neurophysiology,
Vol 115, Issue 10, 2004.

[2] Baccala LA and Sameshima K. {\it Partial directed coherence: a
new concept in neural structure determination}. Biological
Cybernetics, Vol 84, No 6, 2001.

[3] L.~Wu, P.~Neskovic, E.~Reyes, E.~Festa, and W.~Heindel, {\it
Classifying n-back EEG data using entropy and mutual information
features}, European Symposium on Artificial Neural Networks, 2007.

[4] L.~Tian, D.~Erdogmus, A.~Adama, and M.~Pavel, {\it Feature
selection by independent component analysis and mutual information
maximization in EEG signal classification}, IEEE International Joint
Conference on Neural Network, 2005.

[5] J.~Hausser and K.~Strimmer, {\it Entropy inference and the
James-Stein estimator, with application to nonlinear gene
association networks}, Journal of Machine Learning Research, Vol 10,
No 12, 2009.

[6] T.~Yan, J.~Tang, A.~Gong and W.~Wang , {\it Classifying EEG
Signals Based HMM-AR}, International Conference on Bioinformatics
and Biomedical Engineering, 2008.

[7] P.~Ferrez and J.~ Millan, {\it EEG-Based Brain-Computer
Interaction: Improved Accuracy by Automatic Single-Trial Error
Detection}, Advances in Neural Information Processing System, 2007.

[8] L.~Wu and P.~Neskovic, {\it Feature extraction for EEG
classification: representing electrode outputs as a Markov
stochastic process}, European Symposium on Artificial Neural
Networks, 2007.

[9] W.~Xu, C.~Guan, C.~Siong, S.~Ranganatha, M.~Thulasidas and
J.~Wu, {\it High Accuracy Classification of EEG Signal}, in
International Conference on Pattern Recognition, 2004.

[10] A.~Rao, A.~Hero, D.~States, and J~.Engel, {\it Using Directed
Information to Build Biologically Relevant Influence Networks}, in
Journal on Bioinformatics and Computational Biology, Vol. 6, No.3,
June 2008.

[11] B.~Blankertz, G.~Curio, and K~.Muller, {\it Classifying single
trial EEG: Towards brain computer interfacing}, Advances in Neural
Information Processing System, 2002.

[12] G.~W. Oelhert, {\it A note on the Delta method}, The American
Statistician, Vol. 46, No 1, 1992.

[13] D.~Garrett, D.~Peterson, C.~Anderson and M.~Thaut, {\it
Comparison of linear, nonlinear, and feature selection methods for
EEG signal classification}, Vol 11, No 2, IEEE Transactions on
Neural Systems and Rehabilitation Engineering, 2003.

[14] W.~Xu, J.~Wu, Z.~Huang and C.~Guan, {\it Kernel Based Hidden
Markov Model With Applications to EEG Signal Classification}, The
IASTED Int. Conf. on Biomedical Engineering, 2005.

[15] M.~Krumin and S.~Shoham. {\it Multivariate autoregress modeling
and granger causality analysis of multiple spike trains}.
Computational Intelligence and Neuroscience, 2010.

[16] J.~Massey. {\it Causality, feedback and directed information.
Symp Information Theory and Its Applications (ISITA)}, 1990.

[17] Project BCI - EEG motor activity dataset, {\it
http://sites.google.com/site/projectbci/}, Brain Computer Interface
research at NUST Pakistan.

[18] J.~Cohen, W.~Peristein, T.~Braver, L.~Nystrom, D.~Noll,
J.~Jonldes and E.~Smith, {\it Temporal dynamics of brain activation
during a memory task}, Nature, Vol 386, 1997.

[19] H.~Meleis1, S.~Takashima1 and R.~Harner, {\it Analysis of
inphase interaction pattern in EEG}, Bioelectromagnetics, Volume 3,
Issue 1, 1982.

[20] H.~Hui, D.~Pantazis, S.~Bressler, and R.~Leahy, {\it
Identifying True Cortical Interactions in MEG using the Nulling
Beamformer}, Neuroimage, 49(4), 2010.

[21] K.~Ang, Z.~Chin, H.~Zhang, and C.~Guan, {\it Mutual
information-based selection of optimal spatial-temporal patterns for
single-trial EEG-based BCIs}, Pattern Recognition, 2011.

[22] L.~Almeida, {\it MISEP - Linear and nonlinear ICA based on
mutual information}, Journal of Machine Learning Research, 4, 2004.

[23] W.~Hesse, E.~Moller, M.~Arnold and B.~Schack, {\it The use of
time-variant EEG Granger causality for inspecting directed
interdependencies of neural assemblies}, Journal of Neuroscience
Methods, 124(1), 2003.

[24] A.~Delorme, Makeig and S.~EEGLAB, {\it an open source toolbox
for analysis of single-trial EEG dynamics}, Journal of Neuroscience
Methods, 134:9-21, 2004.

[25] http://start.eegspectrum.com/Newsletter/.

[26]  Quinn, C.J. and Coleman, T.P. and Kiyavash, N. and
Hatsopoulos, N.G, {\it Estimating the directed information to infer
causal relationships in ensemble neural spike train recordings},
Journal of computational neuroscience, 30(1), 2011, Springer.

[27] Supp, G.G. and Schlogl, A. and Trujillo-Barreto, N. and Muller,
M.M. and Gruber, T., {\it Directed cortical information flow during
human object recognition: analyzing induced EEG gamma-band responses
in brain's source space}, PLoS One, 2(8), 2007.

[28] Hinrichs, H. and Noesselt, T. and Heinze, H.J., {\it Directed
information flow\--A model free measure to analyze causal
interactions in event related EEG-MEG-experiments}, Human brain
mapping, 29(2), 2008.

[29] Babiloni, F. and Cincotti, F. and Babiloni, C. and Carducci, F.
and Mattia, D. and Astolfi, L. and Basilisco, A. and Rossini, PM and
others, {\it Estimation of the cortical functional connectivity with
the multimodal integration of high-resolution EEG and fMRI data by
directed transfer function}, Neuroimage, 24(1), 2005.

[30] Astolfi, L. and Cincotti, F. and Mattia, D. and Marciani, M.G.
and Baccala, L.A. and de Vico Fallani, F. and Salinari, S. and
Ursino, M. and Zavaglia, M. and Ding, L. and others, {\it Comparison
of different cortical connectivity estimators for high-resolution
EEG recordings}, Human brain mapping, 28(2), Wiley Online Library,
2007.

[31] Korzeniewska, A. and Maczak, M. and Kamiski, M. and Blinowska,
K.J. and Kasicki, S., {\it Determination of information flow
direction among brain structures by a modified directed transfer
function (dDTF) method}, Journal of neuroscience methods, 125(1-2),
Elsevier, 2003.

[32] Lotte, F. and Congedo, M. and L{\'e}cuyer, A. and Lamarche, F.
and Arnaldi, B., {\it A review of classification algorithms for
EEG-based brain--computer interfaces}, Journal of neural
engineering, 4, IOP Publishing, 2007.

[33] Benjamini, Y. and Hochberg, Y., {\it Controlling the false
discovery rate: A practical and powerful approach to multiple
testing}, J. Roy. Statist. Soc. Ser. B. 1995; 57:289-300.

[34] Schfer, J., and K. Strimmer, {\it An empirical Bayes approach
to inferring large-scale gene association net- works},
Bioinformatics,Oct 2004.

[35] Y.~ Benjamini and D.~ Yekutieli, {\it The control of the false
discovery rate in multiple testing under dependency}, The Annals of
Statistics, Vol. 29, No. 4, 1165-1188, 2001.

[36] J.P.~Romano, A.M.~Shaikh and M.~Wolf, {\it Control of the false
discovery rate under dependence using the bootstrap and
subsampling}, TEST, 17, 2008.

[37] X.~Chen, A.~Hero and S.~Savarese, {\it Multimodal Video
Indexing and Retrieval using directed information}, IEEE
Transactions on Multimedia Special Issue on Object and Event
Classification in Large-Scale Video Collections, to appear, Aug,
2011.

[38] H.~Cai, S.~Kulkarni, and S.~Verd¡äu, {\it Universal entropy
estimation via block sorting}. IEEE Transactions on Information
Theory, 50(7):1551-1561, 2004.

[39] J.~Ziv, and A.~Lempel, {\it A Universal Algorithm for
Sequential Data Compression}. IEEE Transactions on Information
Theory, 23(3):337-343, 1977.

[40] J.~Pillow, J.~Shlens, L.~Paninski, A.~Sher, A.~Litke,
E.~Chichilnisky and E.~Simoncelli. {\it Spatio-temporal correlations
and visual signaling in a complete neuronal population}. Nature 454:
995-999, 2008

[41] B.~Krishnapuram, L.~Carin, M.~Figueiredo and A.~Hartemink, {\it
Sparse Multinomial Logistic Regression: Fast Algorithms and
Generalization Bounds}, IEEE Transactions on Pattern Analysis and
Machine Intelligence, 2005.

[42] A.~H.Shoeb, {\it Application of Machine Learning to Epileptic
Seizure Onset Detection and Treatment}, MIT, PhD Thesis, 2009.

[43] Brodmann K. Vergleichende Lokalisationslehre der
Grosshirnrinde. Leipzig : Johann Ambrosius Bart, 1909.




\end{document}